\documentclass[a4paper,fleqn,usenatbib]{mnras}

\usepackage{newtxtext,newtxmath}

\usepackage[T1]{fontenc}
\usepackage{ae,aecompl}

\usepackage{graphicx}	
\usepackage{amsmath}	
\usepackage{amssymb}	

\title{CRTS J035010.7+323230, a new eclipsing polar in the cataclysmic variable period gap}

\author[Mason]{
Paul A. Mason$^{1,2}$\thanks{E-mail: pmason@nmsu.edu}, 
Natalie K. Wells$^{1,2}$,
Mokhine Motsoaledi$^{3,4}$,
Paula Szkody$^{5}$
\newauthor 
Emmanuel Gonzalez$^{6}$\\
$^{1}$New Mexico State University, MSC 3DA, Box 30001, Las Cruces, NM, USA, 88003\\
$^{2}$Picture Rocks Observatory, 1025 S. Solano Suite D., Las Cruces, NM, USA, 88001\\
$^{3}$Department of Astronomy, University of Cape Town, Private Bag X3, Rondelbosch 7701, South Africa\\
$^{4}$South African Astronomical Observatory, P.O. Box 9, Observatory, 7935, South Africa \\
$^{5}$Department of Astronomy, University of Washington, Seattle, WA, USA, 98195\\
$^{6}$Department of Physics, University of Texas at El Paso, El Paso, TX,  USA, 79968\\\
}

\date{Accepted XXX. Received YYY; in original form ZZZ}

\pubyear{2018}

\begin{document}
\label{firstpage}
\pagerange{\pageref{firstpage}--\pageref{lastpage}}
\maketitle{}

\begin{abstract}
We report the discovery of a new eclipsing polar, CRTS J035010.7+323230 (hereafter CRTS J0350+3232). We identified this cataclysmic variable (CV) candidate as a possible polar from its multi-year Catalina Real-Time Transient Survey (CRTS) optical light curve. Photometric monitoring of 22 eclipses in 2015 and 2017 was performed with the 2.1-m Otto Struve Telescope at McDonald Observatory. We derive an unambiguous high-precision ephemeris. Strong evidence that CRTS J0350+3232 is a polar comes from optical spectroscopy obtained over a complete orbital cycle using the Apache Point Observatory 3.5-m telescope. High velocity Balmer and He II $\lambda$4686{\AA} emission line equivalent width ratios, structures, and variations are typical of polars and are modulated at the same period, 2.37-hrs (142.3-min), as the eclipse to within uncertainties. The spectral energy distribution and luminosity is found to be comparable to that of AM Herculis. Pre-eclipse dips in the light curve show evidence for stream accretion. We derive the following tentative binary and stellar parameters assuming a helium composition white dwarf and a companion mass of 0.2 M$_{\odot}$: inclination i = 74.68$^{o}$ ${\pm}$ 0.03$^{o}$, semi-major axis  a = 0.942 ${\pm}$ 0.024 R$_{\odot}$, and masses and radii of the white dwarf and companion respectively: M$_{1}$ = 0.948 $^{+0.006}_{-0.012}$ M$_{\odot}$,  R$_{1}$ = 0.00830 $^{+0.00012}_{-0.00006}$ R$_{\odot}$,  R$_{2}$ = 0.249 ${\pm}$ 0.002 R$_{\odot}$. As a relatively bright (V $\sim$ 17-19 mag), eclipsing, period-gap polar, CRTS J0350+3232 will remain an important laboratory for the study of accretion and angular momentum evolution in polars.

\end{abstract}




\section{Introduction}
Polars are the highest magnetic field subclass of cataclysmic variable (CV) binaries. They consist of a white dwarf with a magnetic field in the 10-250 MG range. The strong magnetic field prevents the formation of an accretion disk and channels the plasma flow from the Roche-lobe filling companion onto the magnetic pole(s) of the white dwarf. Polars are so-named due to the optical/IR cyclotron radiation they emit as accreting plasma spirals down the magnetic field lines just before impact with the white dwarf. In addition, X-ray emission is produced by bremsstrahlung radiation in a post-shock region above the white dwarf surface and/or thermal radiation from the heated white dwarf polar cap. Finally, emission line radiation is observed as the result of recombination in the accretion stream. Polars are well-known for undergoing transitions between actively accreting high states and inactive or less active low states, see e.g. \cite{Wu 2008}, on timescales as short as a few days. For for a still relevant review of polars, see \cite{Cropper 1990} and \cite{Frank 2002} for a theoretical overview of the broader subject.

About 119 confirmed polars are known thanks to several X-ray (especially ROSAT and XMM) and optical (SDSS) surveys. About 27 of these are eclipsing binaries, where the Roche-lobe filling companion orbits close enough to the line of sight, inclination i  > 72$^{o}$, such that we see an eclipse of the white dwarf and eclipses of the much smaller and brighter accretion region(s) located on or near its surface. Most polars have white dwarfs which rotate synchronously with the orbital period of the binary due to strong coupling between the intense white dwarf magnetic field and the donor companion star.  Six or seven polars, $\sim 5\%$, are confirmed asynchronous rotators (with several additional candidates), where the white dwarf is spinning slightly faster or slower than the binary orbit \citep{Campbell 1999}. See \cite{Pavlenko 2018} for a recent study of V1500 Cyg, the first identified asynchronous polar and the site of a bright nova, Nova Cyg 1975. CVs with weaker magnetic fields and/or significantly higher accretion rates  are called intermediate polars or IPs. They usually have a disk that is magnetically disrupted at its inner edge. See \cite{Ferrario 2015} for a recent review of white dwarf magnetism in CVs.

\subsection{The CV period gap}
Relatively few CVs have been found in the period range between 2 and 3 hours, the so-called CV period-gap. For example, in the Ritter-Kolb Cataclysmic Binaries Catalog \citep[][update 2016]{Ritter 2003}, 31 out of 119, or 26\% of polars are in the period gap, while only 7 out of 78, or 9\% of IPs and about 103 out of 1151, or 9\% of non-magnetic CVs fall within the period gap. For orbital periods longer than $\sim$ 3 hours a combination of magnetic braking (the dominant mechanism) and gravitational radiation drive angular momentum loss. The donor star rotates quickly and thus is expected to be magnetically active. A magnetized wind flows from the companion star out of the system along open magnetic field lines, taking with it some angular momentum. This magnetic braking of the donor star rotation rate leads to the loss of total angular momentum from the system. Since the donor rotation is forced to remain tidally synchronized with the binary period, angular momentum lost from the donor's wind is exchanged with the binary and thus the binary semi-major axis decreases. In this way, CVs are expected to evolve from post-common envelope binaries into long period CVs and then into shorter periods CVs, while the donor maintains Roche-lobe contact.

What then can account for the period-gap? A process called disrupted magnetic braking has been proposed \citep{Verbunt 1981, Rappaport 1983} as follows. As the Roche-lobe mass transfer slowly removes mass from the donor, its thermal time-scale increases and thermal relaxation can no longer adjust to the mass transfer. As a result, the companion becomes bloated as compared to the thermal equilibrium case. The star becomes progressively out of equilibrium as the orbital period decreases to near 3 hours. At this point, the companion falls to a mass low enough to become fully convective. As a result, the donor star shrinks and loses contact with its Roche-lobe and then the binary becomes detached. Subsequently in this disrupted magnetic braking scenario, the detached binary evolves slowly towards shorter periods by gravitational radiation alone. Near an orbital period of 2 hours, the secondary refills its Roche-lobe and mass transfer begins again. A prediction of this model is that there must be a large number of difficult to detect detached CVs in the 2-3 hour period range. However, as mentioned, polars as a sub-class have managed to populate this region of the period distribution. A likely explanation for having polars in the period gap is that in polars magnetic braking is less efficient. This is because synchronous rotation allows for the connection between the magnetic field lines of the white dwarf and the donor. Thus there are fewer open field lines available for the escaping wind \citep{Li 1994, Webbink 2002}. Indeed, the idea that the evolution of polars is driven mostly by gravitational radiation as opposed to angular momentum loss due to an escaping magnetized wind from the companion \citep{Wickramasinghe 1994} is becoming more evident.
Eclipsing period-gap polars are important laboratories for the study of CV evolution. So, each new system merits detailed investigation.

In this paper, we present photometry and spectroscopy of a new actively accreting polar in the period-gap, thus further reducing the statistical case for a period-gap for polars. In particular, the presence of both high and low state photometry allows simultaneous eclipse mapping of the white dwarf and the accretion spot independently, assuming a mass-radius relation for the white dwarf and a specific donor mass based on the binary orbital period from \cite{Knigge 2011}.

\section{Observations}

\begin{figure}
\centering
	\includegraphics[width=3.4in]{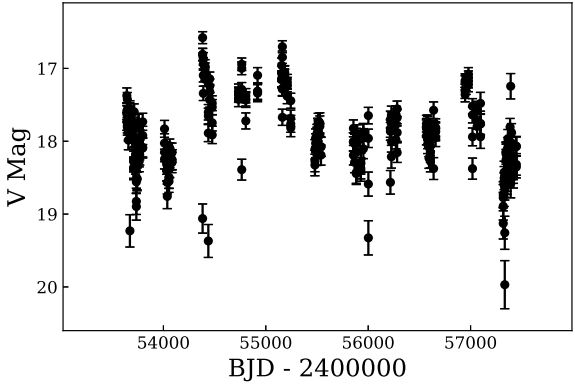}	
	\caption{Catalina Real-Time Survey V-band photometry from late 2005 until early 2016. The roughly two magnitude variation seen in individual seasons shows the orbital variability including occasional eclipse points. The number of faint points shown is consistent with most of them occurring during an eclipse. Notice also that CRTS J0350+3232 undergoes long-term variations in mean brightness of about one magnitude, attributed to a variable accretion rate.}
	\label{Figure1}
\end{figure}

CRTS 035010.7+323230 (hereafter CRTS J0350+3232) was identified as a CV by \cite{Drake 2014} from the Catalina Real-Time Transit Survey (CRTS). The ongoing survey was designed to detect Near-Earth Objects, see \cite{Larson 1998}. It uses the 0.7-m Catalina Schmidt telescope near Tucson, Arizona, providing an 8 deg$^2$ field. Light curves are easily accessible online. We undertook a systematic inspection of candidate CVs in order to identify potential polars by finding CVs that have sudden changes in brightness by several magnitudes. These could indicate transitions between high and low states that are typical of polars. CRTS J0350+3232 was selected for follow-up observations based on those criteria. The second CRTS data release was supplemented with additional, more recent, data graciously supplied to us by Andrew Drake. It includes 290 observations of CRTS J0350+3232 consisting of 30-sec integrations using a V filter. These data were obtained seasonally from 2005 to 2016 and are shown in Figure \ref{Figure1}. Within a given season, the $\sim$ 2 mag variation is consistent with the eclipse of the white dwarf, while the change in average brightness from year to year is attributed to variable accretion rate.

\begin{table}
	\centering
	\caption{Journal of Observations}
	\label{Table1}
	\begin{tabular}{llccr} 
		\hline
		Type & UT Date & $\rm BJD_{\rm TDB}$ - 2450000 & Duration & N\\
		    &(start) &(start)&(hrs)&  \\
		\hline
		Phot & 2015 Dec 8 & 7364.722 & 3.00 & 986\\
		Phot & 2015 Dec 9 & 7365.616 & 4.30 & 1525\\
		Phot & 2015 Dec 10 & 7366.635 & 4.41 & 1589\\
		Phot & 2015 Dec 11 & 7367.650 & 1.50 & 527\\
 		Spec & 2016 Feb 11 & 7429.617 & 2.50 & 10\\
		Phot & 2017 Jan 27 & 7780.569 & 4.29 & 1486\\
		Phot & 2017 Jan 29 & 7782.580 & 4.21 & 1517\\
		Phot & 2017 Jan 30 & 7783.564 & 3.91 & 1410\\
	        	Phot & 2017 Jan 31 & 7784.563 & 4.43 & 1594\\
		Phot & 2017 Feb 1 & 7785.567 & 3.28 & 1182\\
		Phot & 2017 Feb 2 & 7786.564 & 5.20 & 1873\\
		Phot & 2017 Nov 20 & 8077.621 & 7.01 & 8420\\
		\hline
	\end{tabular}
\end{table}


\subsection{Photometry}
High-speed photometry was obtained with the McDonald Observatory 2.1-m Otto Struve Telescope during two weeks, one in December 2015 and one in January/February of 2017, with an additional night in November of 2017. A total of 45.5 hours of quality data were obtained on 11 nights. See Table \ref{Table1} for the Journal of Observations. A broadband, BVR, filter and a ProEm CCD camera were used with 10-sec exposures during most nights and 3-sec exposures during 20 November 2017, when the source was significantly brighter.

\begin{figure}
\centering
	\includegraphics[width=3.3in]{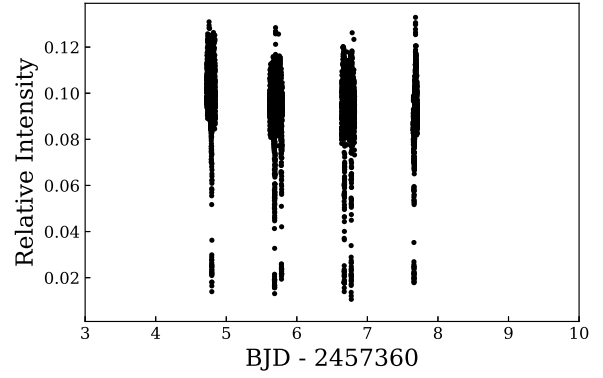}	
	\includegraphics[width=3.3in]{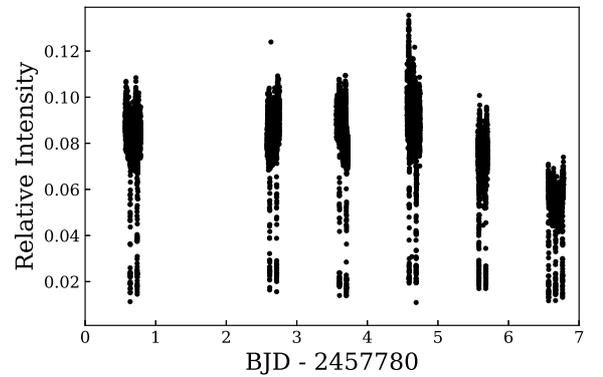}	
	\caption{Light curves obtained with the McDonald Observatory 2.1-m telescope in December 2015 (top) and in January/February 2017 (bottom). Three additional eclipses were observed in November 2017, see Figure \ref{Figure3}.  Typically, two eclipses were observed each night, for a grand total of 22 eclipses.}
	\label{Figure2}
\end{figure}

Science frames were dark subtracted and flat-field corrected using calibration images obtained the same night. Aperture photometry was performed using IRAF \footnote{IRAF is distributed by the National Association of Universities for Research in Astronomy, under cooperative agreement with the National Science Foundation.}, with an aperture size of $\sim$6 arc-seconds. Sky subtraction was performed using a narrow annular region surrounding the aperture. All data, including CRTS photometry, were converted to Barycentric Julian Date in the Barycentric Dynamical Time standard \footnote{ $\rm BJD_{\rm TDB}$ is an absolute time standard which includes the UTC to Terrestrial Time (TT) correction,
 TT = UT + $\Delta$T + 32.184s, where $\Delta$T is the number of leap seconds that have been introduced to UTC.} ($\rm BJD_{\rm TDB}$) using the \cite{Eastman 2010} algorithm. Ten of the eleven nights of McDonald photometry are shown in Figure \ref{Figure2}, where the eclipses are clearly seen. CRTS J0350+3232 light curves are highly variable from night to night, in both average intensity and in shape. Notice the drop in brightness seen during the last two nights of Figure \ref{Figure2}. Even cycle-to-cycle variations are evident.

\begin{figure}
\centering
	\includegraphics[width=3.3in]{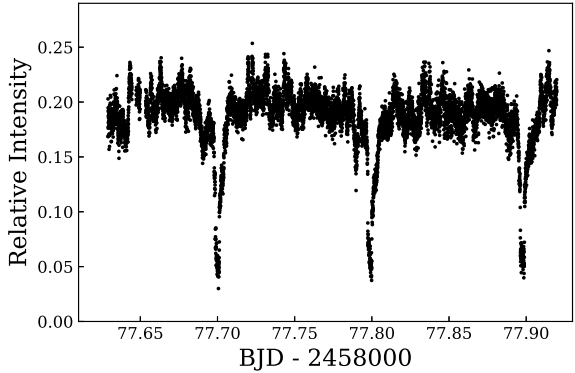}		
	\caption{Light curve obtained 20 November 2017, showing three eclipses with pre-eclipse dips. The light curve on this night is relatively flat other than these features and is brighter than all other nights observed. The pre-eclipse dip is attributed here and in other polars as being due to an occultation of the compact accretion region by the accretion stream.}
	\label{Figure3}
\end{figure}

Only one night of data was obtained in November 2017, but it was by far the longest observation and included three eclipses, see Figure \ref{Figure3}.  On this night, the source was roughly a factor of two brighter; notice the change in the vertical scale between Figures \ref{Figure2} and \ref{Figure3}. These data are about the same brightness, V${\approx}$17 mag, as the brightest level seen in the CRTS photometry, i.e. near BJD = 2455000. In addition, a pre-eclipse dip in intensity occurs in the light curve. This absorption feature is not seen in the light curves at lower brightness levels. An analysis of pre-eclipse dips is given in Section \ref{sec:discussion:dips}.


\subsection{Spectroscopy}
\begin{figure}
	\includegraphics[width=3.1in, clip=true]{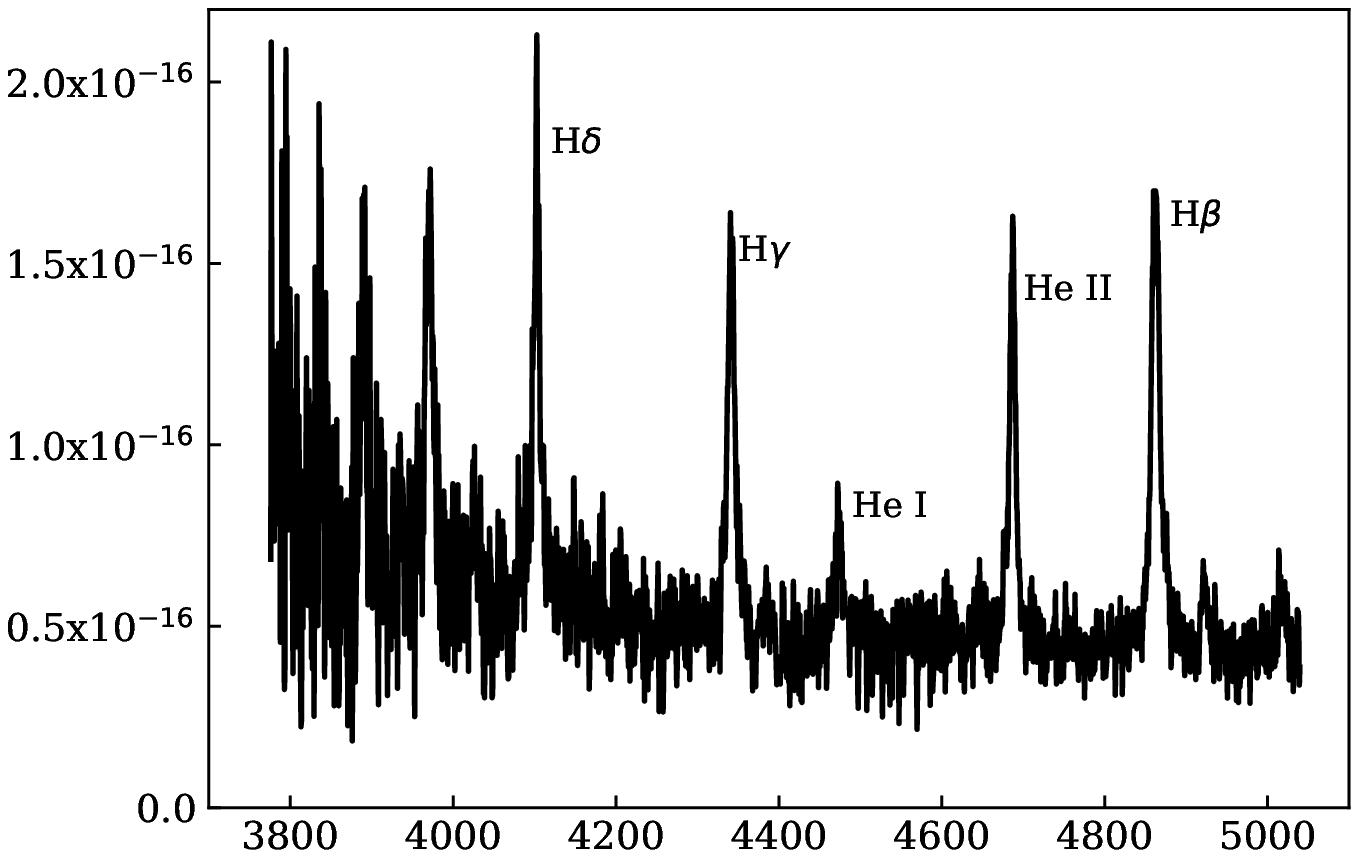}
	\includegraphics[width=3.1in, clip=true]{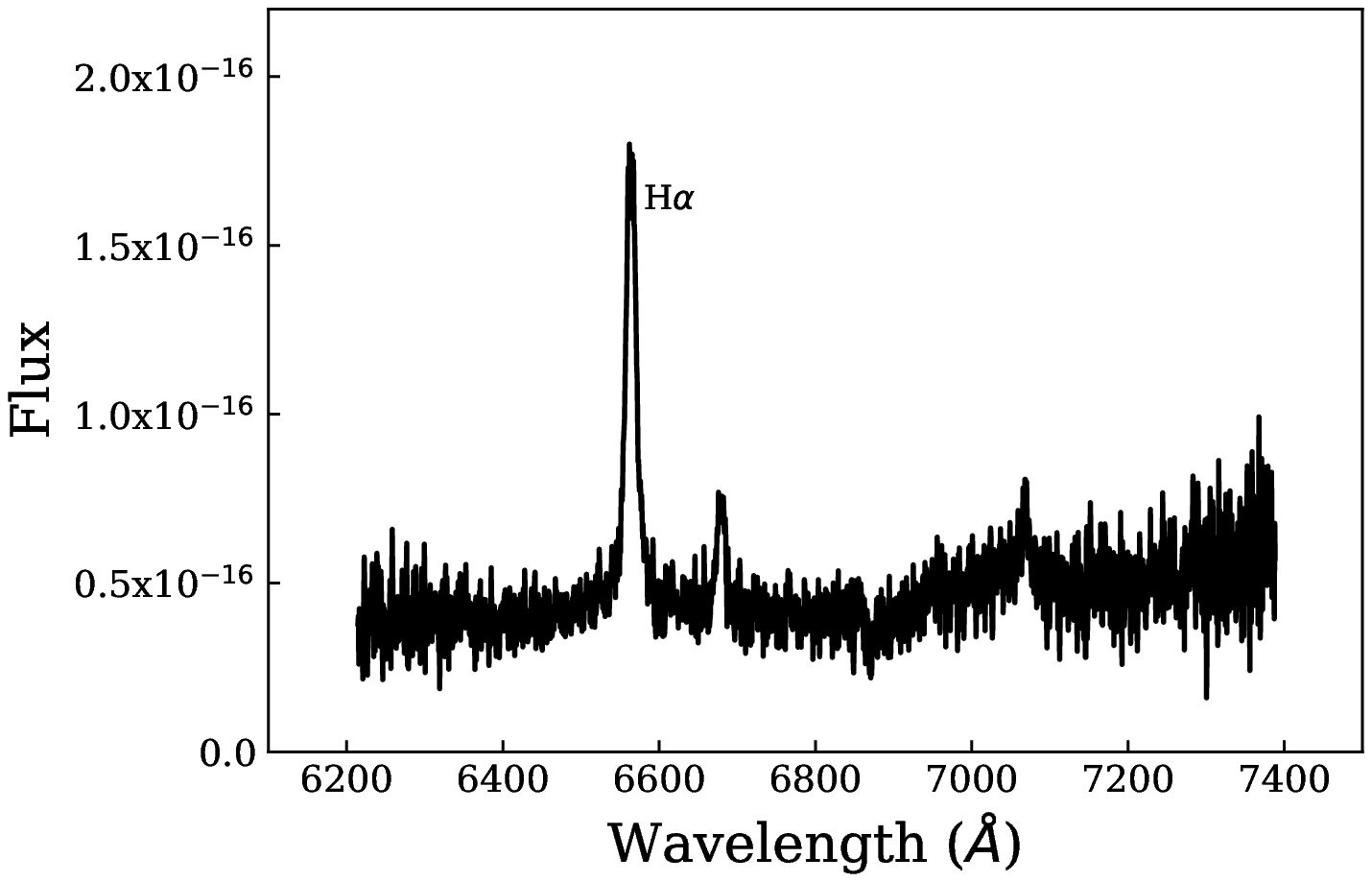}
	\caption{The sum of all spectra of CRTS J0350+3232 with blue and red channels shown in top and bottom panels respectively. Spectra were obtained using the Apache Point Observatory 3.5-m telescope. After a single 10-min integration, nine 15-min integrations were taken over the course of one orbital cycle. The strong hydrogen and helium emission lines typical of polars are present. The broad feature seen near 7075\AA\ is likely a cyclotron emission feature, see text for details.}
	\label{Figure4}
\end{figure}

Spectra of CRTS J0350+3232 were obtained using the Double Imaging Spectrograph on the Apache Point Observatory 3.5-m telescope. Simultaneous blue and red spectra were collected over a continuous 2.5 hours on 11 February 2016. The high resolution gratings were used in each channel, yielding wavelength coverage from 4000 - 5000\AA\ in the blue and 6000 - 7200\AA\ in the red with a spectral resolution of $\sim$0.6\AA\ pixel$^{-1}$. A helium/neon/argon lamp provided wavelength calibrations throughout the time frame, taken about every 30 minutes. The lamp calibration fit was better than 0.01{\AA}. The white dwarf G191B2B was used as a spectrophotometric flux standard. One 10-min spectrum was followed by nine 15-min spectra. Mid-times for all spectra were converted to $\rm BJD_{\rm TDB}$ and then to orbital phase using Equation \ref{eq:Ephem}, derived in the next section, for further analysis. The sum of all spectra is shown in Figure \ref{Figure4}.

\section{Photometric Time-Series Analysis}

\begin{figure}
\centering
	\includegraphics[width=3.3in]{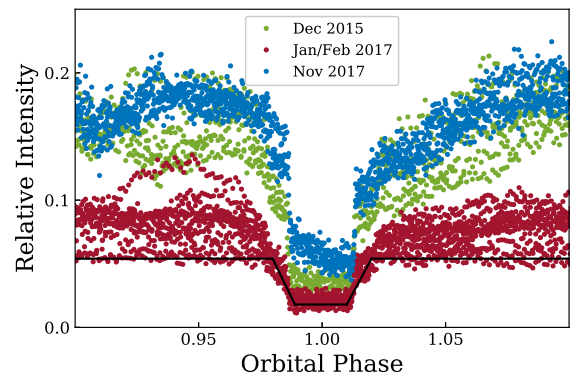}	
	\caption{Individual eclipse profiles from McDonald 2.1-m telescope photometry. The data are phased with the ephemeris, see Equation \ref{eq:Ephem}, derived herein. Notice that when the system is brighter, the eclipse does not appear total. This is interpreted as the visibility of an uneclipsed bright accretion stream. The faintest red curve, showing virtually no active accretion on 2 February 2017, was used in order to model the eclipse of the white dwarf. The model eclipse light curve is shown as a solid black line.}
	\label{Figure5}
\end{figure}

A total of 22 eclipses were observed, recall Figures \ref{Figure2} and \ref{Figure3}. The system was observed in low, intermediate, and high brightness states, see Figure \ref{Figure5}. Such behavior is common in polars. All but 3 eclipses, which occurred on 2 February 2017 when the system was at its lowest observed brightness, display two prominent eclipse features. At the start of each eclipse, a gradual dimming occurs over 2 to 3 minutes and is interpreted as the secondary beginning to block our view of the body of the white dwarf, known as first contact. Then, a distinct and much more rapid drop on the order of 3 to 10 seconds occurs as the bright accretion region on the white dwarf is eclipsed, before the white dwarf is fully eclipsed, which is second contact. Each eclipse dims the system by about 0.6 mag. Corresponding brightenings are seen as first the white dwarf (third contact) and then the bright spot(s) come out of eclipse in turn. The end of the white dwarf partial eclipse is called fourth contact. The sharp features of the bright spot eclipse allow for a straightforward determination of timings for these events. Start of ingress, end of egress, and resulting duration times for each eclipse of the accretion spot were measured and are tabulated in Table \ref{tab:t2_table}.

\begin{table}
	\centering
	\caption{Accretion Region Eclipse Timings}
	\label{tab:t2_table}
	\begin{tabular}{ccccc} 
		\hline
		\multicolumn{2}{c}{\bf Ingress} & \multicolumn{2}{c}{\bf Egress} & {\bf Duration}  \\
		{$\rm BJD_{\rm TDB}$ } & Phase & {$\rm BJD_{\rm TDB}$ } & Phase & (minutes) \\
		{-2450000} & & {-2450000} & & \\
		\hline
		7364.7943 & 0.988 & 7364.7971 & 0.016 & 4.00  \\
		7365.6835 & 0.986 & 7365.6864 & 0.015 & 4.16  \\
		7365.7824 & 0.986 & 7365.7853 & 0.015 & 4.16  \\
		7366.6718 & 0.986 & 7366.6747 & 0.015 & 4.18  \\
		7366.7706 & 0.986 & 7366.7734 & 0.014 & 4.00  \\
		7367.6601 & 0.987 & 7367.6629 & 0.015 & 4.00  \\
		7780.6386 & 0.987 & 7780.6412 & 0.014 & 3.86  \\
		7780.7373 & 0.986 & 7780.7401 & 0.015 & 4.00  \\
		7782.6149 & 0.986 & 7782.6177 & 0.014 & 4.00  \\
		7782.7137 & 0.986 & 7782.7165 & 0.014 & 3.99  \\
		7783.6032 & 0.986 & 7783.6060 & 0.014 & 3.99  \\
		7783.7020 & 0.986 & 7783.7048 & 0.014 & 4.00  \\
		7784.5914 & 0.986 & 7784.5941 & 0.014 & 3.99  \\
		7784.6903 & 0.987 & 7784.6930 & 0.014 & 3.83  \\
		7785.5796 & 0.986 & 7785.5825 & 0.015 & 4.16  \\
		7785.6784 & 0.986 & 7785.6812 & 0.014 & 4.00  \\
		8077.6985 & 0.988 & 8077.7013 & 0.016 & 4.00  \\
		8077.7974 & 0.987 & 8077.8000 & 0.015 & 3.84  \\
		8077.8961 & 0.987 & 8077.8988 & 0.014 & 3.90  \\
		\hline
	\end{tabular}
	\label{Table2}
\end{table}

The rapid changes in brightness at ingress and egress indicate that a compact accretion region is going out of and coming back into view. No evidence of a disc surrounding the white dwarf is apparent in the light curve, as the eclipse would appear broad and deep. For example, see Figure \ref{Figure3}. Identifying times of ingress and egress for the more gradual eclipse of the body of the white dwarf is problematic owing to the nature of the variable un-eclipsed light curve. Sometimes the light curve is increasing in brightness at the start or end of the eclipse, while at other times the opposite is true. In other words, the out-of-eclipse light curve does not appear to be strongly modulated with the orbital period. This is atypical of eclipsing polars and is addressed in the next subsection. However, on the faintest night, when CRTS J0350+3232 was on average $\sim$ 0.5 mag dimmer, three eclipses were observed whose light curves do not show the sharp accretion spot feature and appear much more symmetrical. In this low state we are given a clearer view of the eclipse of the white dwarf, as interference from the out of eclipse variation is minimized. An eclipse from this night of 2 February 2017, showing minimal active accretion, was used to model the eclipse of the white dwarf, see Section \ref{sec:discussion:model}, and is used as the zero-point for the ephemeris derived from this data, see Equation \ref{eq:Ephem}.

\begin{figure}
\centering
	\includegraphics[width=3.3in]{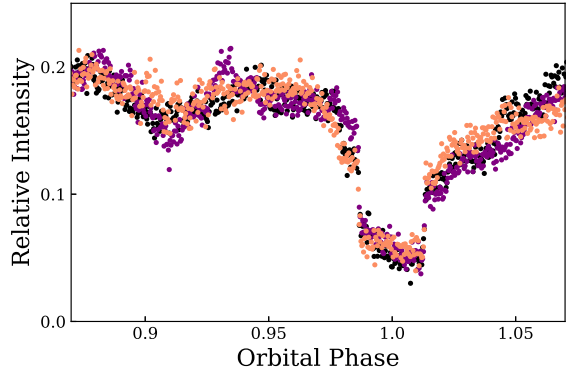}		
	\caption{Three consecutive eclipses observed on 20 November 2017 are phase-folded with the best-fit period. Each eclipse is shown in a different color. A pre-eclipse dip is seen before each eclipse at phase 0.91. Pre-eclipse dips were only seen on this particular night, when the system was at its highest observed brightness. These dips are interpreted as an optically thick stream absorbing light emitted from the accretion zone on the white dwarf.}
	\label{Figure6}
\end{figure}

Compare the shape and brightness of the accretion spot eclipse profiles during high and intermediate state observations to the more symmetric, flat-bottomed eclipses of the low state January/February 2017 data in Figure \ref{Figure5}. Notice that emission from the accretion stream is often in view at mid-eclipse. This is especially evident in the November 2017 data, when the system was at its brightest, see Figure \ref{Figure6}, but is also seen in the December 2015 data during an intermediate state.

\subsection{An Orbital Ephemeris}

\begin{figure}
	\includegraphics[width=3.3in, clip=true]{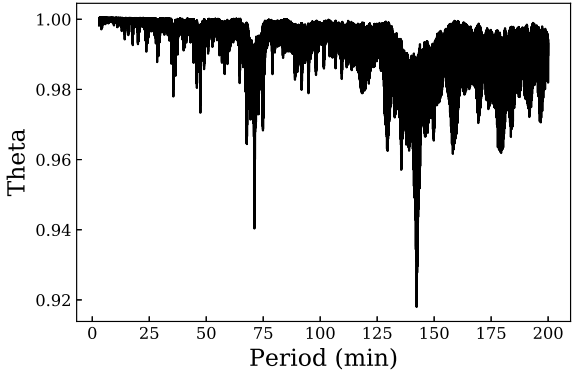}
	\caption{Result of a PDM analysis of the McDonald 2015/2017 data, from 3 to 200-min. The strongest signal, P=142.3042-min (2.37-hrs), agrees well with the orbital period derived from the best fit of the eclipse profiles in the phased McDonald light curve, P=142.3041-min. The secondary signal at 71.15-min represents a harmonic at half the orbital period.}
	\label{Figure7}
\end{figure}

The orbital period is found to high precision by first phasing the eclipse profiles of the 2015 data and the 2017 data separately, basing the fit on the sharp ingress and egress of the compact accretion region. The resulting periods are essentially identical and allow the phasing of the combined 2015/2017 light curves without cycle count ambiguity. This analysis yields an orbital period of P$_{orb}$ = 142.30414(4) min = 2.371736 hr. In addition to this bootstrapping method, a Phase Dispersion Minimization (PDM) analysis \citep{Stellingwerf 1978} of the 2015/2017 data was performed using PyAstronomy, yielding P$_{orb}$ = 142.3042(1) min, see Figure \ref{Figure7}. The PDM method agrees well with eclipse derived period. However, the PDM does not deliver a period that is as precise as the eclipse method. This is not unexpected, as the PDM minimizes dispersion of the entire light curve and thus suffers from dilution from the highly variable out-of-eclipse light curve. In this case, because of the long time separation between the 2015 and 2017 observations, the periodogram shows alias dips surrounding the absolute minimum, however our eclipse phasing analysis allows unambiguous alias exclusion and high precision period determination. 

No significant period is detected in the PDM other than the orbital period and its harmonics. The out of eclipse light curve is variable and aperiodic. Searches for periods using only uneclipsed data also found no significant period other than the orbital period. We interpret the out-of-eclipse light curve as due to a circumpolar accretion spot that is never self-eclipsed by the white dwarf. Our model verified the location of the spot as being consistent with an emission region that never disappears from view thus providing only a low amplitude, weak periodic signature.

Because the accretion spot is a nearly point-like source of light and provides sharp eclipse features, the mutual phasing of all 22 eclipses allows for a unique and precise orbital period determination. It is expressly assumed that the spot is perfectly stable on the surface of the white dwarf; however, using the best of the three low state eclipses (2 February 2017) which more clearly show the ingress and egress of the white dwarf, we anchor the zero-point of the ephemeris to the mid-eclipse of the white dwarf primary. The time of this event is found by taking the average of the ingress and egress times. This means the mid-eclipse of the white dwarf is phase 0, when the secondary is at inferior conjunction. With the mid-eclipse of the white dwarf derived from the McDonald data as phase 0, we obtain the best-fit ephemeris,

\begin{equation}
$$ \rm BJD_{\rm TBD} = 2458077.89751(6) + 0.09882232(3) E. $$
\label{eq:Ephem}
\end{equation}

The precision of this eclipse ephemeris, $\sigma$ = 0.024-s will allow coherent phasing of all types of data for a decade or more. In particular, the accumulation of precise eclipse timings will allow tracking of the orbital evolution of this period-gap polar.

Phasing of the eclipse profiles with the ephemeris is shown in Figures \ref{Figure5} and \ref{Figure6}. There is no discernible trend in phase of ingress or egress (recall Table \ref{Table2}). Measurement residuals are of order 0.001 phase or about 8 seconds, roughly the same as the integration times. The durations of the eclipses of the accretion spot are remarkably stable. A stable accretion structure is expected in mCVs, especially in synchronized polars. Uncertainties of ingress and egress times are 10-sec, yielding an uncertainty in the duration of the accretion spot eclipse of about 14-sec. On a few occasions, the measurement of ingress and egress was more uncertain due to the absence of a clear, sharp drop in brightness. Eclipse durations are more precise for the last three eclipses observed (20 November 2017) due to shorter (3-sec) integration times.

We attempted to further refine the orbital period by including V-band photometry from the CRTS, recall Figure \ref{Figure1}. However, given the paucity of verified eclipse points in the CRTS data due to the ambiguity between low state points and eclipse points, long (30-sec) CRTS integration times, and possible contamination by the stream emission during eclipse, it is not possible to reliably improve P$_{orb}$ using CRTS data.

Using the phase-folded high-time resolution eclipse profiles, the average duration of the bright accretion spot eclipse is 4.00-min (Table \ref{Table2}). Given the orbital period of 142.30-min (2.37-hrs), the half-angle of the eclipse is 5.1$^{o}$. The inclination is a function of mass ratio and this measured eclipse half-angle and is detailed in the discussion.

There is a distinctive difference in the individual eclipse profiles, shown in Figure \ref{Figure5}, between high and low brightness levels. At higher luminosity, the eclipse is not total. This is interpreted as the appearance of the magnetically channeled stream, which causes the flow trajectory to move out of the orbital plane and hence is not completely eclipsed. On the other hand, the pre-eclipse dip (see Figure \ref{Figure6}) which occurs at phase 0.91 only when the system is bright, is most likely due to an optically thick stream absorbing light from the accretion spot on the white dwarf. Most likely, when the luminosity is lower, this part of the stream is less dense and becomes optically thin.

\subsection{Pre-eclipse dip - absorption by the accretion stream}
\label{sec:discussion:dips}

The light curve from a single night in November 2017, when CRTS J0350+3232 was at its brightest, shows three consecutive eclipses with a pre-eclipse dip near phase 0.91. This feature is attributed to absorption by matter flowing in the accretion stream between the two stars. The dips are apparent in Figure \ref{Figure3} and in the phase-folded Figure \ref{Figure6}. For each of the three eclipses, start and end phases, duration, and eclipse depth were measured and are recorded in Table \ref{Table3}. The average depth of the pre-eclipse dip is 0.25 mag. \cite{Katysheva 2012} show that the optical depth of the absorption region is related to the maximum depth in magnitude of the eclipse, given by ${\tau}$ ${\approx}$ 0.92 $ \times$ $\Delta$m. The depths and profiles of the pre-eclipse dips are consistent with those seen in other eclipsing polars, e.g. HU Aquarii ${\Delta}$m = 0.2 mag \citep{Schwope 2014} and FL Ceti ${\Delta}$m = 1.0-1.5 mag \citep{Mason 2015}. The non-eclipsing polar Master J132104+560957.8 displays dips in its light curves associated with emission-absorption line reversals \citep{Littlefield 2018}. Since pre-eclipse dips are common in polars and indicate the presence of an accretion stream, their absence at low accretion rates and presence at high rates as observed in CRTS J0350+3232 is strong support for our classification of CRTS J0350+3232 as a polar.

\begin{table}
	\centering
	\caption{Pre-Eclipse Dip Timings}
	\label{Table3}
	\begin{tabular}{cccccc} 
		\hline
		Mid-Eclipse & Start &  End & Duration & Eclipse & Optical\\
                     ($\rm BJD_{\rm TDB}$ &&&&Depth&Depth\\
		-2450000) & (phase) & (phase) & (min) & (${\Delta}$ M) & (${\tau}$)  \\
		\hline
		8077.6906(3) & 0.877(3) & 0.948(3) & 10.1(4) & 0.22(3) & 0.20\\
 		8077.7897(3) & 0.884(3) & 0.934(3) & 7.1(4)  & 0.36(3) & 0.33\\
		8077.8886(4) & 0.883(4) & 0.959(4) & 10.8(5) & 0.18(4) & 0.17\\
		\hline
	\end{tabular}
\end{table}


\section{Spectroscopic Analysis}
\label{sec:spect_analysis}

The presence of strong emission lines with high radial velocities, especially the presence of He II, is strong evidence that CRTS J0350+3232 is indeed a polar. While the Bowen-blend, characteristic of mCVs, is not apparent in individual spectra nor the sum of all spectra (Figure \ref{Figure4}), the source is quite faint and the blend may be near the level of the noise. This spectrum lacks any absorption features, including diffuse interstellar bands. Near 7050{\AA}, a likely cyclotron feature is present. Unfortunately our long wavelength coverage does not extend much beyond that so any estimation of magnetic field strength for CRTS J0350+3232 would be premature, however the implications of an infrared detection by WISE are discussed in Section \ref{sec:SED}.

Several spectra were examined to see if CRTS J0350+3232 meets a test for magnetic CV classification based on the equivalent widths (EWs) of the emission lines. \cite{Silber 1992} finds that EW H${\beta}$ > 20{\AA} and HeII${\lambda}$4686{\AA} / H${\beta}$ > 0.4 is a sufficient but not necessary indication of magnetic activity for CVs in his sample; Figure 3.10. CRTS J0350+3232 passes this test with EW H${\beta}$ = 20-60{\AA} and HeII${\lambda}$4686{\AA} / H${\beta}$ = 0.48-0.74.

\begin{figure}
	\includegraphics[width=3.3in, clip=true]{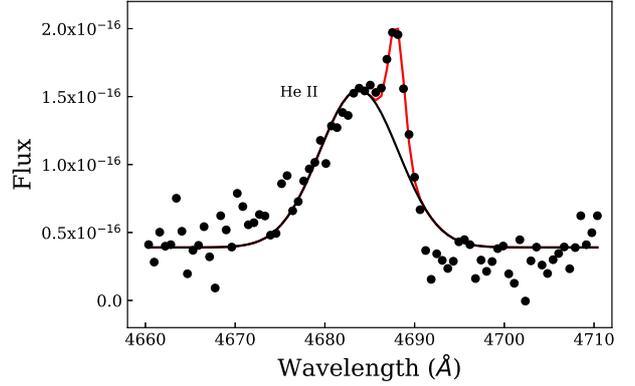}
	\caption{Example double gaussian fit to a He II line. The broad component of the emission line is shown in black, while the narrow component is shown in red. For each line, free parameters include the continuum level, centroid wavelength, height, and width, which are adjusted to minimize ${\chi}^2$ for each component. Centroids are used to calculate radial velocities, and areas under each curve are used to obtain relative intensities for both narrow and broad components. Radial velocity fit parameters are given in Table \ref{Table3}.}
		\label{Figure8}
\end{figure}

In order to study phase-resolved spectra of CRTS J0350+3232, we first note that the strong spectral lines indicate the presence of broad, $\sim$10-15{\AA}, and narrow, $\sim$5{\AA} emission line components. Using a double gaussian fit with ${\chi}^2$ minimization, parameters for FWHM, component velocities, and intensities are measured for the strong Balmer and He lines. An example fit is shown in Figure \ref{Figure8}. The shapes of the strong Balmer and He emission lines are well-fit by a broad emission component and single narrow component which is strongly suggestive of the presence of an accretion stream as opposed to an accretion disk. The eclipse of the white dwarf by the secondary occurred during the first and last spectra of our series. Since the integration times (10-15-min) were much longer than the eclipse, we do not have spectra that allow for the characterization of the donor star.

\begin{figure}
	\includegraphics[width=3.3in, clip=true]{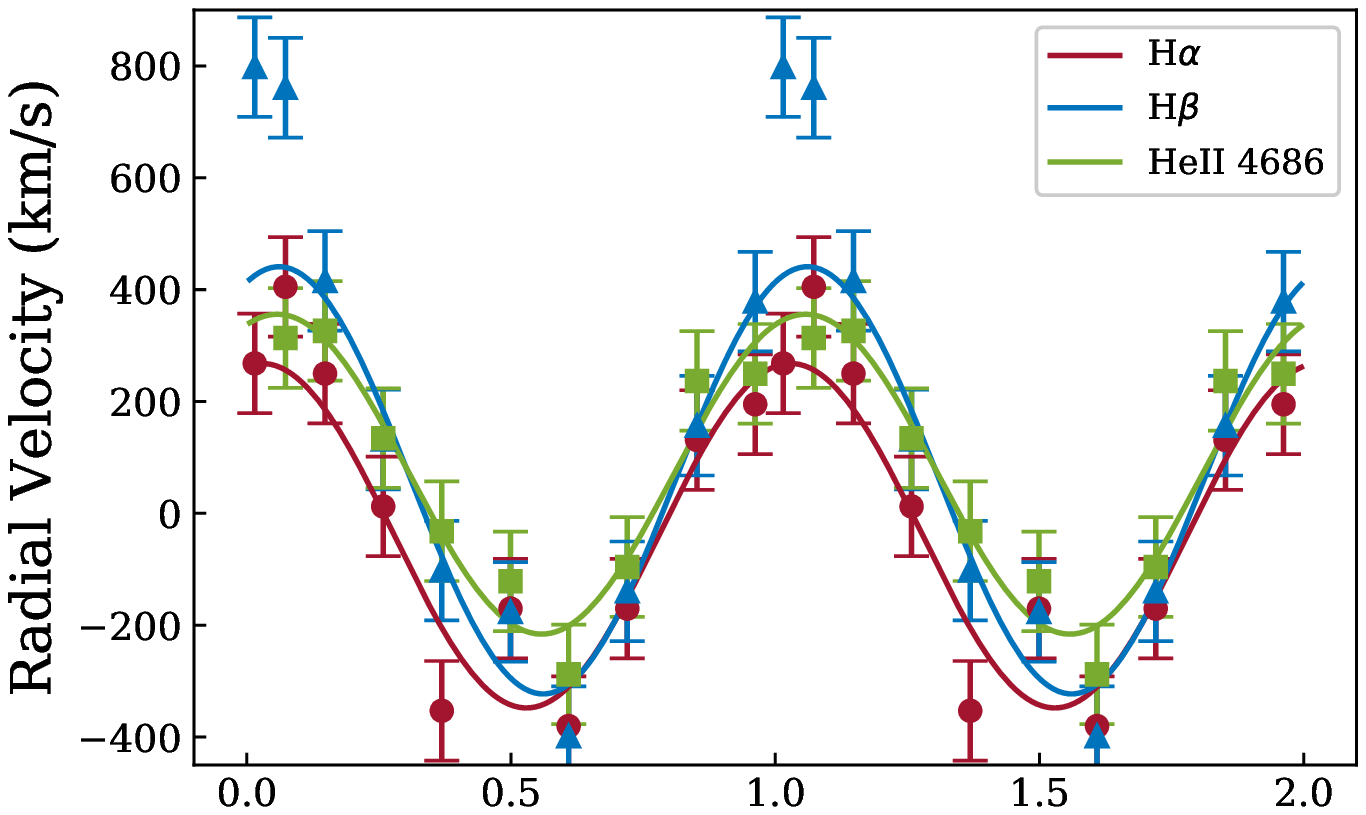}
	\includegraphics[width=3.3in, clip=true]{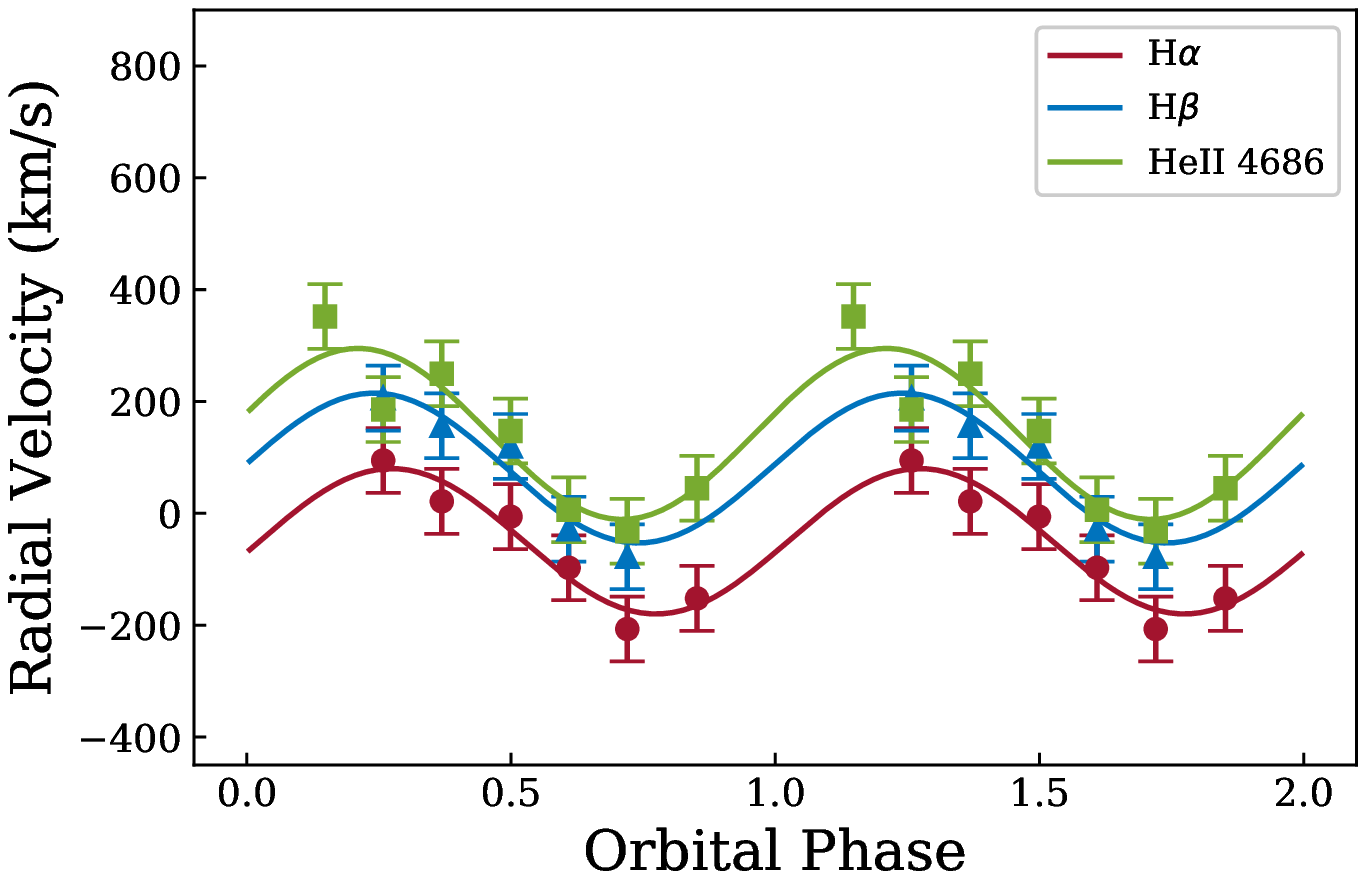}
	\caption{Radial velocity curves for H${\alpha}$, H${\beta}$, and HeII ${\lambda}$4686{\AA} are derived from one orbital cycle of time series spectroscopy of CRTS J0350+3232. The broad and narrow components of each line are shown in the top and bottom panels respectively. The orbital phase is derived from the eclipse of the white dwarf primary as given in the ephemeris. Notice the high velocity points for H${\beta}$ during the eclipse, which are interpreted as high velocity components of an uneclipsed portion of the stream.}
		\label{Figure9}
\end{figure}

\begin{figure}
	\includegraphics[width=3.3in, clip=true]{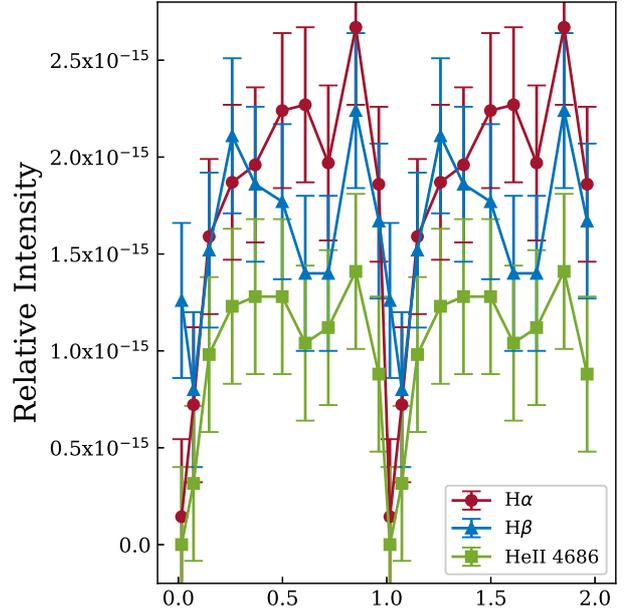}
	\includegraphics[width=3.3in, clip=true]{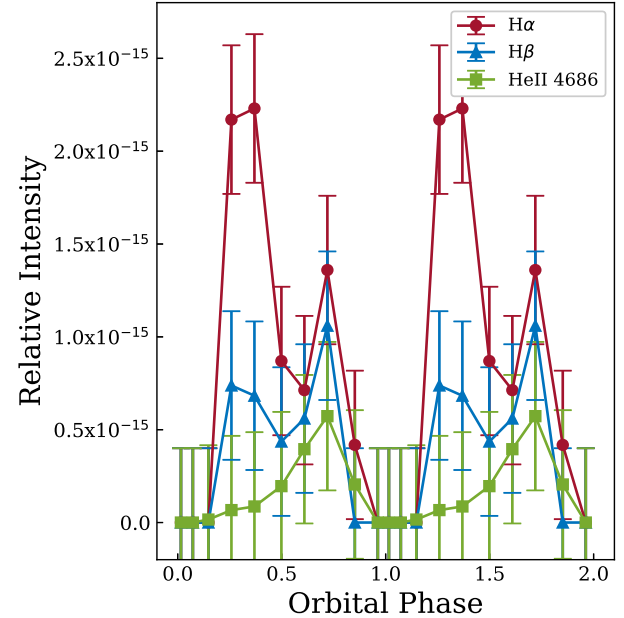}
	\caption{Relative emission line intensities for H${\alpha}$, H${\beta}$, and HeII ${\lambda}$4686{\AA} as a function of orbital phase. The broad and narrow components of each line are shown in the top and bottom panels respectively. The broad components all show similar orbital modulation. For the narrow components, the hydrogen lines show two peaks per orbit whereas the HeII line has only one, indicating that somewhat distinct regions are responsible for this emission.}
		\label{Figure10}
\end{figure}

For each broad and each narrow component, a best-fit to the sinusoidal function:

\begin{equation}
$$V_{R} = \gamma - Ksin({2\pi\over P} + \phi_{RV})$$
\label{eq:RadVel}
\end{equation}
\noindent

was performed. Here, the red-to-blue radial velocity crossing defines 0 spectroscopic phase. For each line, best-fit parameters for the variables in Equation \ref{eq:RadVel} are tabulated in Table \ref{Table4}. In this analysis, P is set to the eclipse period defined by the photometric ephemeris (Equation \ref{eq:Ephem}) while the amplitude K, the radial velocity phase ${\phi_{RV}}$, and the systemic velocity ${\gamma}$, are free parameters in the fit. Best-fit radial velocity curves for the strongest emission lines are shown in Figure \ref{Figure9}, with relative emission line intensities in Figure \ref{Figure10}.

\begin{table}
	\centering
	\caption{Radial Velocity Fit Parameters }
	\label{Table4}
	\begin{tabular}{crrr} 
		\hline
		Line & K (km/s) & ${\phi_{RV}}$ & ${\gamma}$ (km/s)\\
		\hline
		\multicolumn{4}{l}{\bf Broad} \\
		H${\alpha}$ & 308 ${\pm}$ 30 & 0.22 ${\pm}$ 0.01 & -40 ${\pm}$ 20\\
		H${\beta}$ & 382 ${\pm}$ 20 & 0.19 ${\pm}$ 0.01 & 59 ${\pm}$ 15\\
		HeII ${\lambda}$4686\AA & 286 ${\pm}$ 20 & 0.19 ${\pm}$ 0.01 & 70 ${\pm}$ 15 \\	
		HeI & 316 ${\pm}$ 30 & 0.20 ${\pm}$ 0.01 & 204 ${\pm}$ 20 \\
		H${\gamma}$ & 366 ${\pm}$ 30 & 0.17 ${\pm}$ 0.01 & 104 ${\pm}$ 20 \\
		H${\delta}$ & 270 ${\pm}$ 20 & 0.15 ${\pm}$ 0.01 & 60 ${\pm}$ 15 \\
		\multicolumn{4}{l}{\bf Narrow} \\
		H${\alpha}$ & 130 ${\pm}$ 10 & 0.48 ${\pm}$ 0.01 & -50 ${\pm}$ 6\\
		H${\beta}$ & 153 ${\pm}$ 17 & 0.51 ${\pm}$ 0.02 & 136 ${\pm}$ 11\\
		HeII ${\lambda}$4686\AA & 153 ${\pm}$ 15 & 0.54 ${\pm}$ 0.02 & 142 ${\pm}$ 11 \\	
		HeI & 218 ${\pm}$ 15 & 0.64 ${\pm}$ 0.01 & 128 ${\pm}$ 10 \\
		H${\gamma}$ & 205 ${\pm}$ 32 & 0.69 ${\pm}$ 0.02 & 175 ${\pm}$ 22 \\
		H${\delta}$ & 265 ${\pm}$ 22 & 0.65 ${\pm}$ 0.01 & 60 ${\pm}$ 15 \\
		\hline
	\end{tabular}
\end{table}

The photometric eclipse ephemeris (Equation \ref{eq:Ephem}) allows unambigouous phasing of the spectroscopic data. The combined photometric and spectroscopic data are easy to interpret as a typical polar. For example, if there were no accretion column and the emission lines were from the white dwarf, then the emission lines would be oppositely phased with the donor star. So, the red-to-blue radial velocity crossing would occur at the inferior conjunction of the secondary, i.e. mid-eclipse of the white dwarf. In this case, $\phi_{RV}$ = $\phi_{Ecl}$. If rather, the lines arise from an offset area (like an accretion flow and column) then there will be an offset in phase between the lines and the photometry. This is indeed the case and the measured offset is $\phi_{Ecl}$ = $\phi_{RV}$ + 0.251. The combination of offset emission lines decomposed into broad and narrow components, the latter of which track the heated face of the companion, is a common feature of polars and again points to a polar classification for CRTS J0350+3232.

\begin{figure}
\centering
	\includegraphics[width=3.3in]{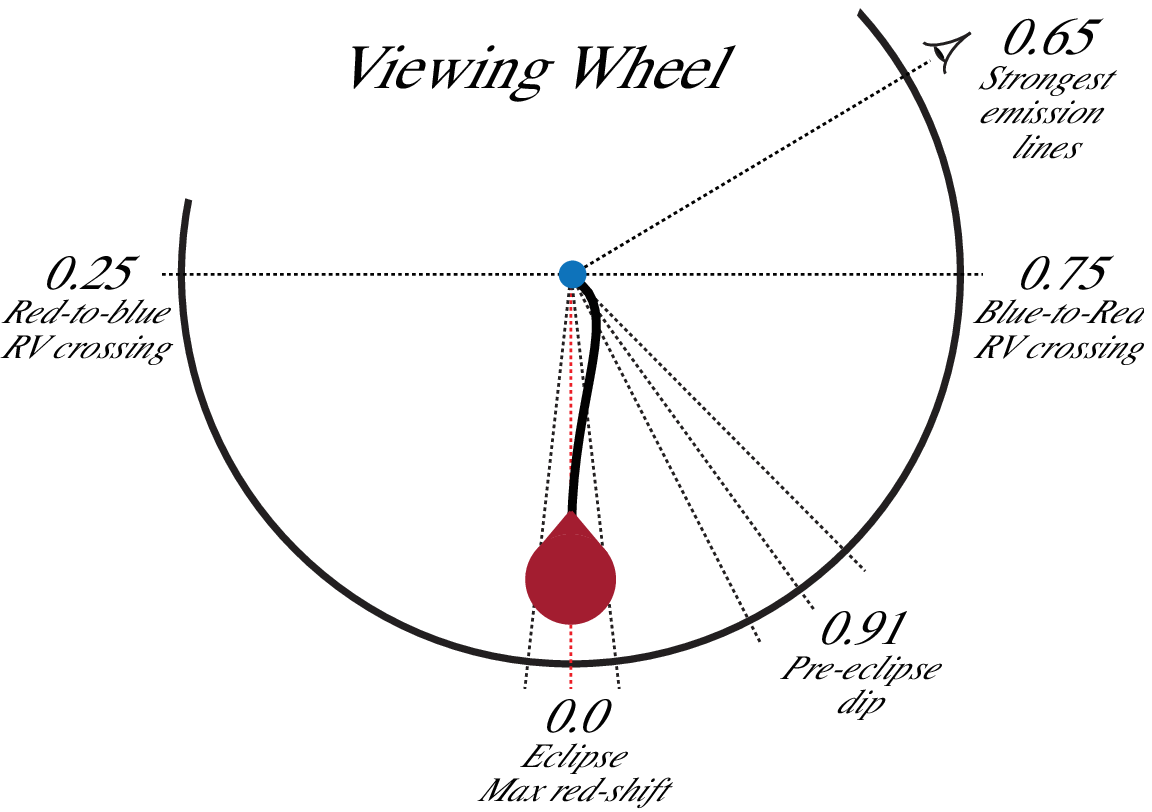}	
	\caption{An illustration of the binary and accretion stream. As the binary orbits, the observer views the accretion flow as a function of phase approximately as shown. The brightest part of the stream is relatively near the white dwarf as evidenced by the very high emission-line velocities. The Roche-lobe filling companion star slightly overlaps the dotted lines showing the eclipse because the inclination is less than 90 degrees. }
	\label{Figure11}
\end{figure}

The observer views the binary as illustrated in Figure \ref{Figure11}. The maximum observed broad component radial velocity occurs during the eclipse. The transitions from blue-to-red and red-to-blue broad component radial velocities occur at eclipse phases 0.751 and 0.251 respectively. The emission-lines are brightest when their view is most favorable with the largest surface area visible, namely eclipse phase 0.65.

\section{Discussion}
\label{sec:discussion}

In this section, we present the spectral energy distribution for this object and compare it to the prototype polar AM Herculis. Then, a combined white dwarf - accretion spot eclipse model is developed and the resulting parameters are given. Finally, pre-eclipse dips are discussed and the maximum optical depth of the stream is estimated.

\subsection{The Spectral Energy Distribution}
\label{sec:SED}

\begin{figure}
\centering
	\includegraphics[width=3.3in]{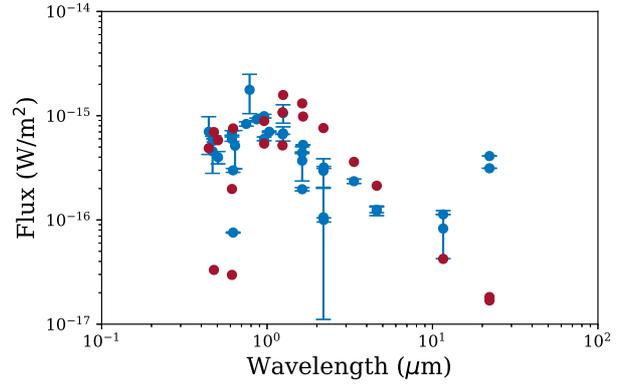}	
	\caption{Spectral energy distribution plot of CRTS J0350+3232 (blue) and the prototype polar AM Herculis (red) obtained with the online Vizier SED tool. AM Herculis points are scaled to the distance (576-pc) of CRTS J0350+3232 using the GAIA DR2 parallax and are corrected for extinction.}
	\label{Figure12}
\end{figure}

In order to further characterize CRTS J0350+3232, its spectral energy distribution (SED) was constructed using data from the online VizieR Photometry viewer. For comparison, we extracted the corresponding data for the prototype polar AM Herculis (AM Her). AM Her fluxes were then scaled to the distance of CRTS J0350+3232 using the GAIA DR2 distances of 88-pc and 576-pc respectively. Fluxes were then corrected for extinction using the reddening function of Schlafly \& Finkbeiner in the NASA/IPAC Infrared Science Archive. The scaled AM Her SED is shown along with CRTS J0350+3232 in Figure \ref{Figure12}. Both VizieR data sets include optical observations from high to low luminosity states. In particular, the three points in the lower left of Figure \ref{Figure12} indicate low states of both systems. The SEDs are very similar, with the exception of the far infrared wavelengths, which are higher in CRTS J0350+3232. A natural explanation for this difference is that CRTS J0350+3232 was in a high state during the WISE W3 and W4 observations and that cyclotron emission is detected, see \cite{Harrison 2015} for a comprehensive study of WISE observations of polars, in which Figure 2 suggests that a W3 detection implies that CRTS J0350+3232 may be a low (10 MG) magnetic field polar. WISE detection provides further support for the classification of CRTS J0350+3232 as an accreting polar in the period gap. 

Only a few points in the CRTS light curve occur during the eclipse. Probably all of these included some stream emission and thus do not represent a true low state, as the donor is likely fainter than the detection limit, m$_V{\approx}$20 mag, of the CRTS telescope. Our McDonald photometry is limited to m$_V{\approx}$21 - 22 mag. Using the Gaia DR2 distance of 576 $^{+125}_{- 88}$pc \citep{Brown 2018} and an absolute magnitude for a period gap CV donor of M$_V{\approx}$12 mag \citep{Knigge 2011}, the donor star's apparent magnitude, corrected for extinction, is estimated as m$_V$${\approx}$22, well below the CRTS limit. The out-of-eclipse intensity is 3.0 times greater than during the eclipse. Given the uncertainties of these estimates, we find that the absolute magnitude of the white dwarf is about 1 magnitude greater than the donor, hence M$_V{\approx}$11 mag.

CRTS J0350+3232 was not detected by ROSAT, and we initiated a director discretionary Swift XRT 2.4-ks observation on 25 July 2016 in collaboration with K. Mukai. Only two X-ray photons in the 0.3 to 10 keV range were detected, which is insignificant. This is not strongly constraining as CRTS J0350+3232 is intrinsically faint and often found in low accretion states, as shown in Figure \ref{Figure1}. 

\subsection{Eclipse modeling}
\label{sec:discussion:model}

In this section, we present a model to constrain 8 parameters: 4 binary parameters, orbital period P, semi-major axis a, eccentricity e, inclination i; and the stellar masses M$_{1}$ and M$_{2}$ and stellar radii R$_{1}$ and R$_{2}$. Standard eclipsing binary light curve modeling also provides a measure of the temperature ratio, however this is bandpass dependent and better determined spectroscopically, thus we do not include this in our model. In the case of CRTS J0350+3232, high state data showing eclipses of the accretion spot, as well as low state observations providing a clear white dwarf eclipse profile relatively uncontaminated by other light sources, along with the selection of the donor mass M$_{2}$ based on the empirical mass-radius relation, provide a complete solution for the binary parameters as well as information about the size and location of the accretion region.

The white dwarf eclipse duration, ingress/egress duration, and period determination provide three measurements in an attempt to solve a system with eight unknowns. Following usual practice the stellar radii are scaled with the semi-major axis, R$_{1}$/a and R$_{2}$/a. See \cite{Parsons 2017} for a discussion of several ways in which the degeneracy between the sizes of the two stars and the inclination of the system may be broken. Since the binary is very close and tidally locked, we assume e = 0, so four additional constraints are required. 

In principle, this degeneracy can be broken using photometry of the secondary eclipse. However, the lowest luminosity state seen in Figure \ref{Figure5} from 2 February 2017 displays a relatively flat out-of-eclipse light curve ($\sim$0.2 mag variation) with no indication of a secondary eclipse. This is not surprising since the white dwarf is much brighter and much smaller than the donor and the donor at m$_{V}$${\approx}$22 mag is near the detection limit of the observations. 

\begin{figure}
\centering
	\includegraphics[width=3.3in]{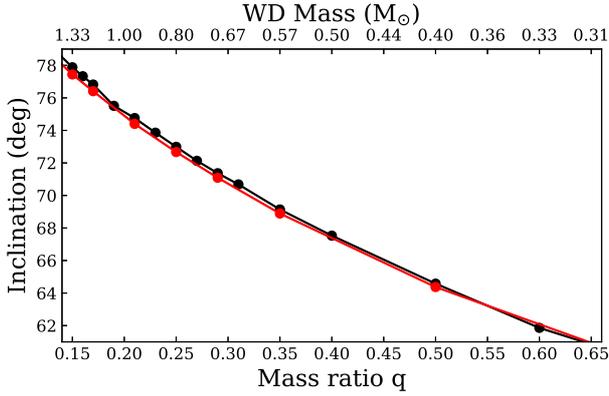}	
	\caption{Inclination versus mass ratio as independently derived from the white dwarf eclipse (black) and the spot eclipse (red) (see Table \ref{Table5}). While the bottom horizontal axis is generally true, the top horizontal axis is defined only for M${_2}$=0.2 M$_{\odot}$.}
	\label{Figure13}
\end{figure}

\begin{figure}
\centering
	\includegraphics[width=3.3in]{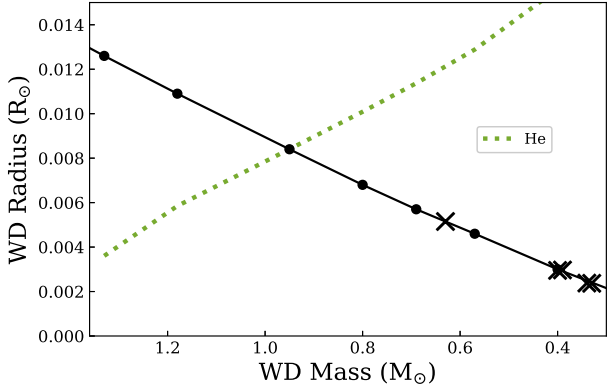}	
	\caption{The white dwarf radius versus mass is shown for eclipse solutions (solid black curve) with radial velocity solutions marked by Xs. The dotted line represents the mass-radius relation for a helium composition white dwarf \citep{Nauenberg 1972}.}
	\label{Figure14}
\end{figure}

Two equations, Newton's version of Kepler's third law, and the Roche-lobe filling condition of the donor \citep{Kopal 1959, Kopal 1972, Eggleton 1983, Knigge 2011}, reduce the number of needed constraints to two. The high precision measurement of white dwarf eclipses obtained during low state and spot eclipses obtained during high state (see Table \ref{Table5}) places a strong constraint on the inclination as a function of mass ratio, q = M$_{2}$/M$_{1}$ \citep{Chanan 1976}, see Figure \ref{Figure13}. Extreme values of the mass ratio allow values from 65$^{o}$< i < 78$^{o}$. However, the strongest constraints are obtained using a combined white dwarf eclipse model and the K$_{2}$ radial velocity measurements. Each narrow spectral component presented in Table \ref{Table4} allows for an independent estimate of the mass ratio via the mass-function. A particular solution is found when the donor mass is set to 0.20 M$_{\odot}$. This selection of donor mass is based on the \cite{Knigge 2011} sequence of model physical parameters for CV donor stars. For period gap polars, he finds M$_{2}$ = 0.20 M$_{\odot}$. Mass function solutions are represented by Xs in Figure \ref{Figure14}. Disentangling the emission line components, including a K-correction due to the center of light offset, resulted in an unexpectedly large range in narrow component values. The result is inconsistent with a pure helium white dwarf. Given this preliminary spectroscopic analysis, we suggest that the white dwarf may be composed of elements significantly heavier than helium. A more detailed spectroscopic study is needed to resolve this issue.

The full list of best fit model parameters assuming a pure helium composition white dwarf is given in Table \ref{Table6}. Note however, that by assuming an accretion spot on the surface of the white dwarf, the model introduces three spot parameters: the spot size and its latitude and longitude on the white dwarf. The combined white dwarf plus spot eclipse model provides three strong constraints on the position and size of the spot. Latitude and longitude are measured from the point on the white dwarf surface facing the donor on the line of centers between the stars and longitude is measured eastward of this point. The longitude of the spot is well constrained by the lag in time between the mid-times of the spot and white dwarf eclipses. A spot longitude of 28${^o}$ indicates that the stream follows a ballistic path which is magnetically channeled fairly quickly, as shown schematically in Figure \ref{Figure11}, and is also consistent with the phase of the pre-eclipse dip. The strongest constraint on the spot latitude comes from the fact that the spot never appears to be self-eclipsed by the white dwarf. Hence, 75$^{o}$ < l < 90$^{o}$ where l is the latitude of the spot on the surface of the white dwarf. The radius of the spot is constrained by the duration of the spot ingress and egress and is found to be 0.0009 R$_{\odot}$ which is about 10\% of the radius of the white dwarf.

Finally, the ellipsoidal variation of the donor is not large enough to be detected in the out-of-eclipse light curve in the lowest luminosity state observed on 2 February 2017. It is important to note that there is also an indication of low-level accretion around the eclipse at phase 0.0. This was taken into consideration in determining contact point times for the white dwarf eclipse by considering the brightness of the system well outside of eclipse. In addition this allows for an estimate of the uncertainty in the eclipse timing based on a low state model with ellipsoidal variations of the donor at a level of $\sim$0.1 mag and a small amount of accretion taking place during this low state. The non-detection of ellipsoidal variations in our light curve is not strongly constraining, however this is consistent with a low temperature donor, which agrees with our assumed mass, our estimated apparent magnitude m$_{V}$${\approx}$22 mag and a relatively low inclination, as we derived.

\begin{figure}
\centering
	\includegraphics[width=3.3in]{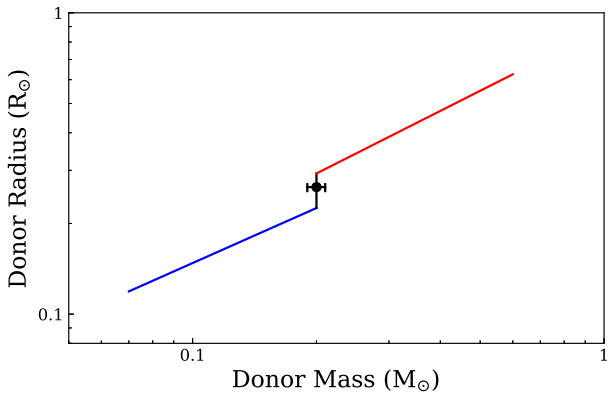}	
	\caption{The mass and radius of CRTS J0350+3232 along with the empirical broken-power-law fit is shown. The blue track is the mass-radius relation for fully convective stars. See Section \ref{sec:discussion:model} for details.}
	\label{Figure15}
\end{figure}

The assumed mass and derived radius of the CRTS J0350+3232 donor star are plotted in Figure \ref{Figure15} as a filled circle. The solid line is a broken-power-law fit representing the semi-empirical evolutionary sequence of CV donor stars from  \cite{Knigge 2011}. The vertical line depicts the reduction in radius which occurs when the donor becomes fully convective after the donor has been reduced to about 0.2 M$_{\odot}$, as expected for CVs with orbital periods of about 3 hours. It is interesting and important to note that the solution does not fall on the fully convective mass-radius relation (lower blue line in Figure \ref{Figure15}). This suggests that the donor is bloated compared to a fully convective counterpart. This bloating is consistent with the fact that the donor in CRTS J0350+3232 is filling its Roche-lobe, as indicated by active accretion, during its time in the period gap. 

In summary, we used both the white dwarf and accretion spot eclipses to derive the inclination (i) as a function of mass ratio (q) (Figure \ref{Figure13}). The independent measurement of both the white dwarf and accretion spot eclipses, combined with the Roche-lobe radius estimation of \cite{Eggleton 1983} as updated by \cite{Knigge 2011}, and Kepler's Third Law are employed. These are used along with our measurements of radial velocity amplitudes of the narrow component emission lines and an assumption of the donor mass (M$_{2}$ = 0.20 M$_{\odot}$ from \cite{Knigge 2011}), to constrain the other parameters: white dwarf mass M$_{1}$, radii R$_{1}$ and R$_{2}$, semi-major axis (a), and inclination (i). The results are given in Table \ref{Table6}. Uncertainties in the derived parameters are from propagation of errors from the best effort solutions, see Figure \ref{Figure14}. We adopt the solution intersecting the green dotted line of Figure \ref{Figure14}, indicating solutions consistent with the pure helium white dwarf mass-radius relation of \cite{Nauenberg 1972}. We find that the donor is bloated compared to a fully convective star of the same mass, which is consistent with a Roche-lobe filling donor despite having a period within the 2-3 hour period gap. A more comprehensive spectroscopic study is needed to resolve the discrepancy between emission line measurements and our tentative solution.


\begin{table}
	\centering
	\caption{Eclipse Contact Points for Model Input}
	\label{Table5}
	\begin{tabular}{ccccc} 
		\hline
		 & \multicolumn{2}{c}{\bf White Dwarf} & \multicolumn{2}{c}{\bf Accretion Region} \\
		 & Phase & Intensity & Phase & Intensity \\
		\hline
		t$_{1}$ & 0.980 & 0.054 & 0.988 & 0.117 \\
		t$_{2}$ & 0.989 & 0.018 & 0.989 & 0.083 \\
		t$_{3}$ & 1.010 & 0.018 & 1.014 & 0.065 \\
		t$_{4}$ & 1.020 & 0.054 & 1.015 & 0.095 \\
		\hline
	\end{tabular}
\end{table}

\begin{table}
	\centering
	\caption{Model Output}
	\label{Table6}
	\begin{tabular}{l} 
		\hline
		 \multicolumn{1}{c}{\bf Binary Parameters}\\
		 P = 142.30414 ${\pm}$ 0.00004 min \\
		 a = 0.942 ${\pm}$ 0.024 R$_{\odot}$\\
		 q = 0.211 ${\pm}$ 0.004\\
 		 i = 74.68$^{o}$ ${\pm}$ 0.03$^{o}$\\ 
		 \multicolumn{1}{c}{\bf Stellar Masses and Radii}\\
		 M$_{1}$ = 0.948 $^{+0.006}_{-0.012}$ M$_{\odot}$ \\ 
	     M$_{2}$ = 0.20 \\
		 R$_{1}$ = 0.00830 $^{+0.00012}_{-0.00006}$ R$_{\odot}$ \\
 		 R$_{2}$ = 0.264 ${\pm}$ 0.002 R$_{\odot}$ \\ 
		 \multicolumn{1}{c}{\bf Spot Parameters}\\
		 Radius = 0.0009 ${\pm}$ 0.0001 R$_{\odot}$\\
		 Longitude = 28$^{o}$ ${\pm}$ 16$^{o}$\\
		 Latitude = 82.5$^{o}$ ${\pm}$ 7.5$^{o}$\\
		\hline
	\end{tabular}
\end{table}

\section{Conclusions}

We present the discovery of CRTS J0350+3232, an actively accreting eclipsing polar within the CV period gap. Eclipses allow for unambiguous identification of the orbital period and over time will allow the study of an important stage in CV period evolution. 

The combination of several results from our photometric and spectroscopic campaign establish CRTS J0350+3232 as a polar. 

1) CRTS J0350+3232 undergoes transitions between high and low luminosity states. The 11-year CRTS light curve shows both short term orbital modulation and long term variability in accretion rate, which is typical of polars.

2) The binary is found to be eclipsing and high speed photometry shows detailed ingress and egress which occur in just a few seconds. These features demonstrate that the accretion zones are very small, a small fraction of the radius of the white dwarf. This is consistent with a magnetically columnated accretion flow.

3) The presence of strong Balmer and HeII${\lambda}$4686$\AA$ emission lines make the source a strong CV candidate and radial velocity curves indicate that the source is a polar. Our radial velocity curves derived for both broad and narrow components show the same period as the eclipse period.
  
4) \cite{Silber 1992} finds that HeII ${\lambda}$4686\AA / H${\beta}$ ratio versus the equivalent width of H${\beta}$ is a good diagnostic for magnetism in CVs. The emission lines of CRTS J0350+3232 meet these criteria for classification as a magnetic CV. 

5) The SED of CRTS J0350+3232 is quite similar to that of the prototypical polar AM Herculis. We scaled multiwavelength observations of AM Herculis to the distance of the source using Gaia parallaxes and found consistency across wavelengths.

6) CRTS J0350+3232 shows evidence of stream accretion. Pre-eclipse dips are best explained as the bright spot being eclipsed by the accretion stream. In addition, high radial velocities for the broad emission component indicate a magnetically columnated flow onto the white dwarf surface. The narrow component indicates that the surface of the donor is irradiated. 

Additional multi-wavelength observations are needed to further constrain the binary and stellar parameters and emission processes of CRTS J0350+3232. More spectroscopic observations are needed to better constrain the spin period of the white dwarf through emission-line radial velocity studies. X-ray and polarimetric observations are needed in order investigate the structure of the bremsstrahlung and cyclotron accretion regions respectively. Being a short period polar, CRTS J0350+3232 is a strong candidate as a radio source, see \cite{Barrett 2017}.


\section*{Acknowledgements}

We thank Andrew Drake for clarifying issues regarding CRTS data, and providing the most recent data on this object. We thank David Buckley, Colin Littlefield, Ed Sion, and Koji Mukai for thoughtful discussions. We thank the anonymous referee for suggestions that greatly benefitted this paper. PM and NW thank the SDSS/FAST program and PM and EG thank the NSF/PAARE program for funding. This work was supported in large part by Picture Rocks Observatory. We thank Thomas Harrison for attempting IR observations of this source and Koji Mukai for a quick-look analysis of a short Swift observation which resulted in a non-detection. PS acknowledges support from NSF grant AST-1514737. Photometry was obtained using the 2.1-m telescope of the McDonald Observatory of the University of Texas. Spectral data were based on observations using the Apache Point 3.5-m telescope, which is owned and operated by the Astrophysical Research Corporation. 

This research has made use of the VizieR catalogue access tool, CDS, Strasbourg, France. The original description of the VizieR service was published in A\&AS 143, 23. This work has also made use of data from the European Space Agency (ESA) mission {\it Gaia} (\url{https://www.cosmos.esa.int/gaia}), processed by the {\it Gaia} Data Processing and Analysis Consortium (DPAC, \url{https://www.cosmos.esa.int/web/gaia/dpac/consortium}). Funding for the DPAC has been provided by national institutions, in particular the institutions participating in the {\it Gaia} Multilateral Agreement.





\bsp	
\label{lastpage}
\end{document}